%% file: main.tex
\documentclass[oribibl]{svproc}
\usepackage{graphicx}

\let\OLDthebibliography\thebibliography
\renewcommand\thebibliography[1]{
	\OLDthebibliography{#1}
	\vspace*{-0.75\baselineskip}
	\setlength{\parskip}{0pt}
	\setlength{\itemsep}{0pt plus 0ex}
}

\usepackage{float}
\usepackage{xcolor}
\usepackage{booktabs}
\usepackage{array}
\usepackage[hidelinks]{hyperref}
\usepackage{enumitem}
\usepackage{amsfonts}
\usepackage{amsmath}
\usepackage{bm}
\usepackage{bbm}

\newcommand{\yvet}{\bm{y}}

\newcommand{\Ivet}{\bm{I}}

\newcommand{\Kvet}{\bm{K}}

\newcommand{\Wvet}{\bm{W}}

\newcommand{\Unovet}{\bm{1}}

\newcommand{\muvet}{\bm{\mu}}
\newcommand{\epsvet}{\bm{\varepsilon}}

\newcommand{\omegavet}{\bm{\omega}}

\newcommand{\Omegavet}{\bm{\Omega}}

\DeclareMathOperator*{\argmin}{arg\,min}

\usepackage[hidelinks]{hyperref}
\usepackage{xurl}

\expandafter\def\expandafter\normalsize\expandafter{%
	\normalsize%
	\setlength\abovedisplayskip{4pt}%
	\setlength\belowdisplayskip{4pt}%
	\setlength\abovedisplayshortskip{4pt}%
	\setlength\belowdisplayshortskip{4pt}%
}

\begin{document}
\mainmatter              
\title{Energy load forecasting using Terna public data: \\a free lunch multi-task combination approach}
%

\author{Daniele Girolimetto\thanks{E-mail address: \email{daniele.girolimetto@unipd.it}} \and Tommaso Di Fonzo}
\institute{Department of Statistical Sciences, University of Padova, \\
	Via C. Battisti 241, 35121 Padova, Italy}

\maketitle  

\begin{abstract}
We propose a quick-and-simple procedure to augment the accuracy of 15-minutes Italian load forecasts disaggregated by bidding zones published by Terna, the operator of the Italian electricity system. We show that a stacked-regression multi-task combination approach using Terna and daily random walk na\"{i}ve forecasts, is able to produce significantly more accurate forecasts immediately after Terna publishes on its data portal the energy load measurements for the previous day, and the forecasts for the current day.
\end{abstract}

\section{Italian energy load forecasting by Terna}

The short-term forecasting of electricity demand plays a vital role in the power grid management.
Given the deployment of intermittent renewable energy sources and the ever increasing consumption of electricity,
the generation of accurate demand-side electricity forecasts is very valuable to grid operators,
to optimize resource use, grid stability, and reliability.
In this context, load forecasting is critical for predicting future energy demand and managing the intermittency and variability of renewable energy resources to ensure a stable performance.
The control and scheduling of the demand for electricity using high frequency time series forecasting is a powerful methodology used in power distribution systems worldwide and has become of primary importance for energy suppliers in nowadays competitive electricity markets.
The literature is rich with forecasting models for country load forecasting.
However, an ongoing challenge is the development of a broadly applicable method for electricity forecasting across geographic
locations (e.g., bidding zones).

\begin{figure}[!tb]
	\centering
	\includegraphics[width = \linewidth]{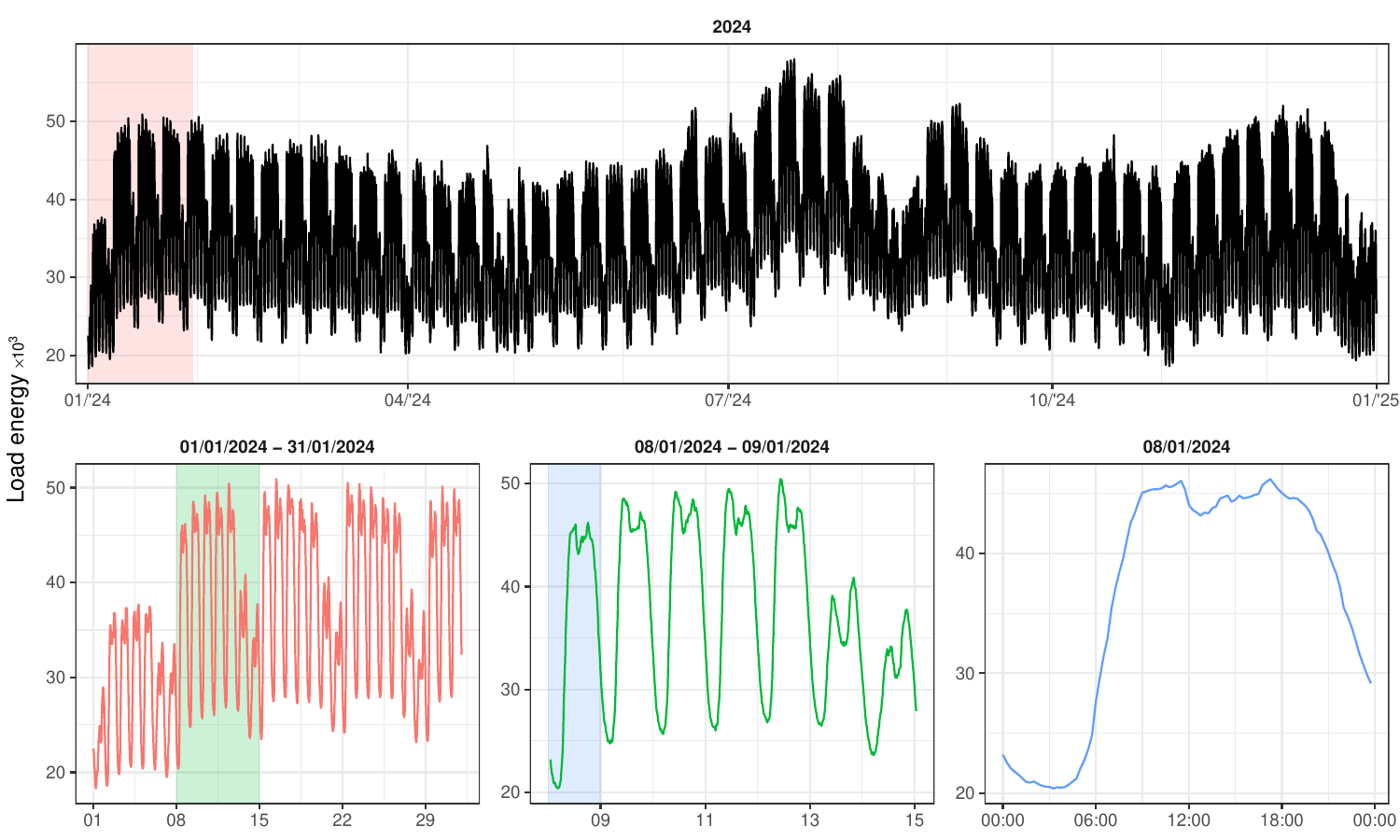}
	\vspace*{-2em}
	\caption{15-minutes time series of the Italian energy demand in 2024 (top panel). The bottom panels show the time series of January 2024 (left), the second January week (center), and January 8, 2024, respectively.
	}
	\label{fig:ts_italy}
	\vspace*{-1em}
\end{figure}

Terna, the Italian state grid company, has the task of ensuring the continuity and security of the electricity supply\footnote{Terna is the Europe's largest independent electricity Transmission System Operator (TSO). It processes the official statistics of the entire national electricity sector and is responsible for official communications to international bodies such as Eurostat, IEA, OECD, UN.}. At the very heart of all Terna does is dispatching: an essential, strategic process carried out at every moment throughout the day, in order to guarantee a balance between electricity supply and demand throughout the Italian grid. Among the various activities carried out for this purpose, Terna currently publishes on its data portal short-term load forecasts for the next day, at national level and disaggregated by seven bidding zones (North, Centre-North, Centre-South, South, Calabria, Sicily and Sardinia). The time frequency is every 15 minutes, i.e., for each zone the forecast load values at 96 different time points during the day are available. In addition, on the same portal historical 15-minutes time series of observed (see for example \autoref{fig:ts_italy}) and forecast load as of year 2020 may be easily downloaded by anyone interested\footnote{At \url{https://dati.terna.it/en/} one may find the graphs of the real-time Total Load of all bidding zones for the current day, and of the 96 Forecast Load for the entire current day. The numerical values of both series are shown on the graphs, and can be easily collected. Furthermore, at  \url{https://dati.terna.it/en/download-center\#/load/total-load} the 15-minutes time series of actual and forecast load for all bidding zones up to the end of the previous day are publicly available.}. This is a wealth of information, and we want to start evaluating the potential influence in Italian electricity market of this recent data-access opportunity.

In this paper, a quick-and-simple strategy is proposed to improve the market operator's forecasts of very short-term Italian electricity demand, which is of interest for power retailers in managing dispatching and bid strategies. No further forecasting model is proposed, rather we obtain 15-minutes Italian energy demand forecasts disaggregated according to billing zones through a stacked-regression \cite{Breiman1996} post-processing approach using only the public information (observed data and forecast records) made available by Terna.
The methodology is validated via out-of-sample comparisons using power load data from December 2023 to December 2024 to train the approach and evaluate its performance.
We show that significant improvement of the accuracy of the original Terna forecasts may be achieved quickly and with relatively little effort.

\section{Multi-task forecast combination}

Let $\yvet \equiv \yvet_{t+h}= \begin{bmatrix} y_{1,t+h} & \ldots & y_{i,t+h} & \ldots & y_{n,t+h} \end{bmatrix}^\top \in \mathbb{R}^{n}$ be the vector of the realizations of the $n$-variate time series $\yvet_{t}$ we are interested in forecasting, where $t$ and $h \ge 1$ denote the forecast origin and horizon, respectively. We assume that $p \geq 2$ experts independently produced base forecasts for every one of the $n$ individual components of $\yvet$, yielding $m = np$ base forecasts overall\footnote{\cite{DGTDF2025} present a complete discussion of the optimal multi-task combination approach, including (i) possible linear constraints between the variables, and (ii) the general case where the experts may produce base forecasts only for a subset of the totality of variables.}. Denote $\widehat{y}_{i}^j \equiv \widehat{y}_{i,t+h|t}^j$ the unbiased base forecast of the $i$-th individual variable of $\yvet$ produced by the $j$-th expert, and $\varepsilon_{i}^j = \widehat{y}_{i}^j - y_{i,t+h}$ the zero-mean base forecast error. Let $\widehat{\yvet}^j \in \mathbb{R}^{n}$, $j=1,\ldots,p$, be the vector of base forecasts produced by the $j$-th expert, with error given by $\epsvet^j = \widehat{\yvet}^j - \yvet$. Finally, denote $\widehat{\yvet} = \text{vec}\begin{bmatrix} \widehat{\yvet}^1 & \dots & \widehat{\yvet}^j & \dots & \widehat{\yvet}^{p} \end{bmatrix}  \in \mathbb{R}^m$, and $\epsvet = \text{vec}\begin{bmatrix} \epsvet^1 & \dots & \epsvet^j & \dots & \epsvet^{p} \end{bmatrix} \in \mathbb{R}^m $. 
	
Under the assumption that the base forecasts are unbiased, the relationship between the target $\yvet$ and the base forecasts $\widehat{\yvet}$ can be represented through the stacked regression model $\widehat{y}_i^j = y_i + \varepsilon_i^j$, $i=1,\ldots,n$, $j=1,\ldots,p$, that is:
\begin{equation}\label{eq:mod}
\widehat{\yvet} = \Kvet \yvet + \epsvet ,
\end{equation}
where $\Kvet = \left(\Unovet_p \otimes \Ivet_n\right) \in \{0,1\}^{m \times n}$, and $\Wvet =E(\epsvet\epsvet^\top) \in \mathbb{R}^{m \times m}$ is a p.d. error covariance matrix. As we shall see, $\Wvet$ plays a crucial role in determining the base forecasts contribution to the multi-task combined forecasts.
	
For a single variable $y_i$, the combined forecast $\widehat{y}_i^c$ is typically expressed as a weighted linear combination of the $p$ base forecasts for that variable \cite{Bates1969}:
$$
\widehat{y}_i^c = \omegavet_i^\top \widehat{\yvet}_i = \sum_{j=1}^p \omega_{ij} \widehat{y}_i^j, \quad i=1, \ldots, n,
$$
where $\widehat{\yvet}_i =
\left[\widehat{y}_{i}^1 \; \ldots \; \widehat{y}_{i}^j \; \ldots \; \widehat{y}_{i}^p\right]^\top \in \mathbb{R}^p$ and $\omegavet_i \in \mathbb{R}^p$ contains the combination weights, which are often estimated as to minimize the forecast error variance for the single $i$-th variable \cite{Bates1969, Newbold1974}. We extend this principle to the multivariate case so that the combined forecast $\widehat{\yvet}^c \in \mathbb{R}^n$ integrates information across all variables and experts while accounting for the interdependence between variables.
	
Starting from the regression model (\ref{eq:mod}), it can be shown \cite{DGTDF2025} that the minimum mean square error (MMSE) linear combined predictor $\widehat{\yvet}^c$, obtained as solution to the linearly constrained quadratic program
\begin{equation*}
	\widehat{\yvet}^c = \argmin_{\yvet}\left(\widehat{\yvet} - \Kvet\yvet\right)^\top\Wvet^{-1}\left(\widehat{\yvet} - \Kvet\yvet\right) ,
\end{equation*}
is given by
\begin{equation}
	\label{eq:ymtfc}
	\widehat{\yvet}^c = \Omegavet^\top\widehat{\yvet},
\end{equation}
where $\Omegavet = \Wvet^{-1}\Kvet\Wvet_c$, with $\Wvet_c = \left(\Kvet^\top\Wvet^{-1}\Kvet\right)^{-1}$. Denoting $\muvet = E\left(\yvet\right)$, $\widehat{\yvet}^c$ is unbiased, i.e., $E\left(\widehat{\yvet}^c\right) = \muvet$, with error covariance matrix $\Wvet_c$. In addition, $\Wvet_c \preceq \Wvet_{j}$, $j=1,\ldots,p$, where $\Wvet_j = E\left[\epsvet^j\left(\epsvet^j\right)^\top\right] \in \mathbb{R}^{p \times p}$, meaning that the single expert's base forecasts  $\widehat{\yvet}^j$ are always not better (i.e., the error covariance matrix is not `smaller') than the stacked-regression-based  multi-task combined predictor $\widehat{\yvet}^c$. In other terms - assuming that the base forecasts are unbiased and the error covariance matrix $\Wvet$ is known - for a multivariate time series the combined predictor $\widehat{\yvet}^c$ does not reduce the precision of the base forecasts.

\section{Combination of Terna and na\"{i}ve forecasts}

Our proposal is based on the multi-task combination approach presented so far, where the 15-minutes base forecasts produced by only two `experts' are considered (i.e., $p=2$): $\widehat{y}_{i,t+h|t}^{\text{terna}}$ and $\widehat{y}_{i,t+h|t}^{\text{drw}}$. The former is the forecast provided by Terna, which integrates a comprehensive set of influencing factors, including meteorological conditions, climate trends, and socio-economic variables, reflecting a sophisticated approach to electricity load prediction. The latter is a daily random walk (drw) model, $\widehat{y}_{i,t+h|t}^{\text{drw}} = y_{i,t-96+h}$, a naïve forecasting approach where the 15-minutes load forecast for the following day is equal to the corresponding 15-minutes current day’s observed value.
Obviously, the quality of the na\"{i}ve forecasts is nowhere near as good as that of the Terna forecasts. However, we show that globally combining the two approaches across all the variables through a stacked-regression results in a significant accuracy improvement over the Terna forecasts. Put simply, Terna forecasts may benefit from the publicly available information by a minimal-cost and computationally trivial approach like drw.

The forecasting experiment focuses on the 8 Italian energy load time series, spanning from December 4, 2023, to December 31, 2024, with data collected at 15-minute intervals from the Terna download center\footnote{The data were downloaded on January 15, 2024.}. The Italian electricity market has a natural hierarchical structure with a single level, where the national load is disaggregated into seven components: North, Centre-North, Centre-South, South, Calabria, Sicily, and Sardinia. To evaluate the forecasting accuracy of our proposal, we implement a rolling forecast experiment with daily iterations throughout the entirety of 2024 (January 1 to December 31, encompassing 366 days due to the leap year). For each forecast day, we utilize the previous four weeks data as validation set to compute optimal weights and error covariance matrices. Denoting with $k=1$ the \textit{daily random walk} (drw) forecasts, for comparison purposes we consider six combination approaches:
\begin{itemize}[nosep, leftmargin=!, align=left]
	\item[$\bm{k=2,3,4}$.] \textit{Equal weights} (ew) and two \textit{local weighting single-task combination} proposed by \cite{Bates1969} and \cite{Newbold1974}, called respectively lw$_{\text{var}}$ and lw$_{\text{cov}}$.
	\item[$\bm{k=5,6}$.] Two \textit{sequential local-combination-then-reconciliation} (scr$_{\text{var}}$ and scr$_{\text{cov}}$), proposed by \cite{DGTDF2025}.
	\item[$\bm{k=7}$.] \textit{Global weighting multi-task combination} (gw), according to expression (\ref{eq:ymtfc}), where the base forecasts of all zones are taken into account, and $\Wvet$ is estimated by a shrunk covariance matrix where the validation error covariances between different variables and different experts are set to zero.  In order to achieve coherency, Italy forecasts are obtained as the sum of the combined forecasts of the seven bidding zones.
\end{itemize}
The first three approaches are local-single-task combination procedures \cite{Thompson2024} producing incoherent forecasts (i.e., the sum of combined forecasts of the 7 bidding zones does not equal the combined forecast of Italy). The remaining approaches are both global-multi-task and coherent \cite{DGTDF2025}. The forecast accuracy is evaluated using the Geometric Average ($GA$) of the Relative Mean Absolute Error ($RMAE$ \cite{Davydenko2013}), with Terna forecast as benchmark: 
$$
RMAE_{i,h}^{k} =  \frac{MAE_{i,h}^{k}}{MAE_{i,h}^{\text{terna}}}, \; i=1,\ldots,8, \; h=1,\ldots,96, \; k=1,\ldots,7,
$$ 
with $MAE_{i,h}^{k} = \frac{1}{366}\scalebox{.8}{$\displaystyle\sum_{q = 1}^{366}$} \big|y_{i,q+h}-\widetilde{y}_{i,q+h|q}^{\,k}\big|$, where $y_{i,q+h}$ is the actual value at the $h$-th 15-minutes interval of day $q$, and $\widetilde{y}_{i,q+h|q}^{\,k}$ is the combined forecast produced by the $k$-th procedure at the forecast origin $q$. 

From \autoref{tab:res}, it appears that the gw approach consistently outperforms the Terna benchmark and all the other combination procedures, whether local or global, always achieving the lowest $GA\text{-}RMAE$. Specifically, compared to the Terna benchmark, in terms of MAE gw yields improvements ranging from 5.95\% (C-South) to 7.17\% (North), with Italy standing at 7.1\%\footnote{If MSE is used instead of MAE, gw registers even better results, ranging from 10.31\% (C-South) to 12.97\% (North), with Italy at  12.74\%.}.
\begin{table}[t]
	\caption{$GA\text{-}RMAE$ (benchmark: Terna forecasts). Bold entries identify the best performing 
		approaches. Red color denotes a forecasting performance worse than the benchmark. `BZ' stands for `bidding zones'. \label{tab:res}}
		\vspace*{-0.5\baselineskip}
	\resizebox{\linewidth}{!}{
		\input{scoreT_nomse_MAE.tex}}
		\vspace*{-1.5\baselineskip}
\end{table}

\begin{figure}[!t]
	\centering
	\includegraphics[width = 0.95\linewidth]{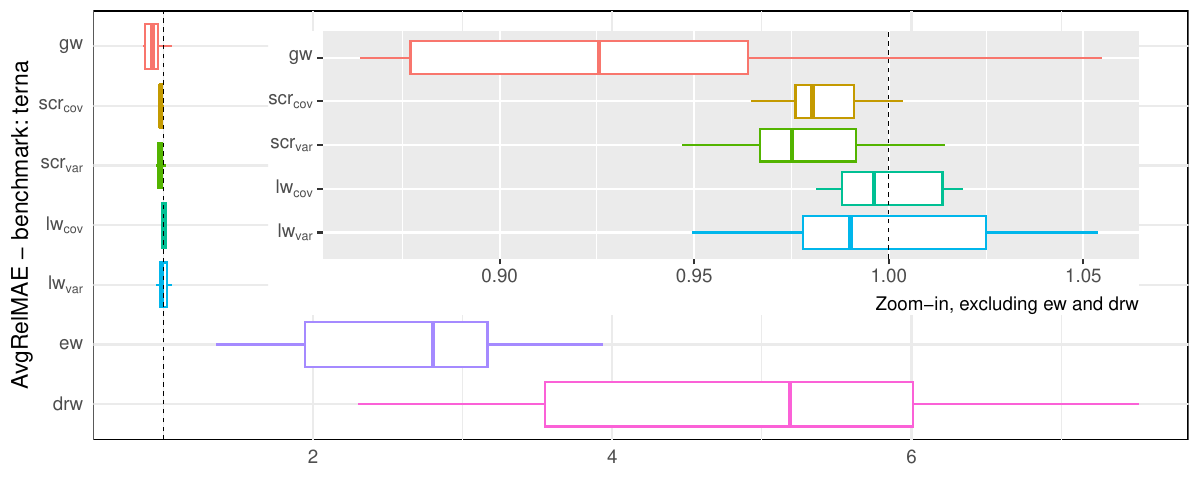}
	\vspace*{-1em}
	\caption{Boxplots of the $RMAE$ values for Italy across the 96 forecast horizons.}
	\label{fig:Terna_boxplot}
	\vspace*{-1em}
\end{figure}

\begin{figure}[H]
	\centering
	\vspace*{-1em}
	\includegraphics[width=0.65\linewidth]{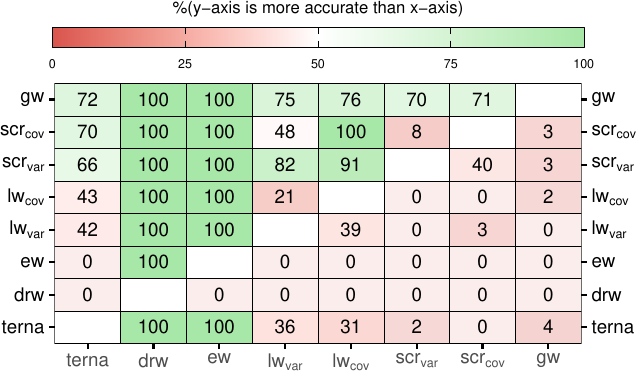}
		\vspace*{-1em}
	\caption{ Pairwise DM-test results for Italy evaluated using absolute loss for $h = 1, \dots, 96$. Each cell reports the percentage of series for which the $p$-value of the DM-test is below $0.05$: e.g., the value 72 in the top-left cell means that gw resulted more accurate ($p$-value $<0.05$) than terna in 72\% of the forecast horizons.\label{fig:dmtest}} 
	\label{fig:italy}
	\vspace*{-2em}
\end{figure}

The $RMAE$ boxplots for Italy in \autoref{fig:Terna_boxplot} provide a clear visualisation of the overall better performance of gw compared to the benchmark Terna across the 96 forecast horizons. Finally, these findings are further supported by the Diebold-Mariano tests reported in \autoref{fig:italy}, showing statistically significant improvements in absolute loss across all horizons, highlighting the methodological advantages of the proposed procedure, in addition to the practical ones given by the simplicity of the required calculations.

\bibliographystyle{abbrv} 

\bibliography{biblio.bib}

\end{document}

%% file: scoreT_nomse_MAE.tex
\begin{tabular}{>{}l|>{\centering\arraybackslash}p{4em}>{\centering\arraybackslash}p{4em}>{\centering\arraybackslash}p{4em}>{\centering\arraybackslash}p{4em}>{\centering\arraybackslash}p{4em}>{\centering\arraybackslash}p{4em}>{\centering\arraybackslash}p{4em}>{\centering\arraybackslash}p{4em}|>{}c|c}
\toprule
\multicolumn{1}{c}{\textbf{}} & \multicolumn{8}{c}{\textbf{Total and zones}} & \multicolumn{1}{c}{\textbf{}} & \multicolumn{1}{c}{\textbf{}} \\
\cmidrule(l{3pt}r{3pt}){2-9}
\textbf{App.} & Italy & North & C-North & C-South & South & Calabria & Sicily & Sardinia & \textbf{BZ} & \textbf{All}\\
\midrule
drw & \textcolor{red}{4.6734} & \textcolor{red}{5.7779} & \textcolor{red}{5.1600} & \textcolor{red}{4.4835} & \textcolor{red}{6.0456} & \textcolor{red}{4.5806} & \textcolor{red}{3.1216} & \textcolor{red}{2.2264} & \textcolor{red}{4.2663} & \textcolor{red}{4.3152}\\
\addlinespace[1ex]
ew & \textcolor{red}{2.5392} & \textcolor{red}{3.0748} & \textcolor{red}{2.7354} & \textcolor{red}{2.3801} & \textcolor{red}{3.1053} & \textcolor{red}{2.4023} & \textcolor{red}{1.7057} & \textcolor{red}{1.2939} & \textcolor{red}{2.2893} & \textcolor{red}{2.3192}\\
lw$_{\text{var}}$ & \textcolor{red}{1.0004} & \textcolor{red}{1.0025} & \textcolor{black}{0.9990} & \textcolor{black}{0.9978} & \textcolor{black}{0.9966} & \textcolor{black}{0.9950} & \textcolor{black}{0.9799} & \textcolor{black}{0.9561} & \textcolor{black}{0.9894} & \textcolor{black}{0.9908}\\
lw$_{\text{cov}}$ & \textcolor{red}{1.0001} & \textcolor{red}{1.0011} & \textcolor{red}{1.0001} & \textcolor{black}{0.9982} & \textcolor{black}{0.9988} & \textcolor{black}{0.9984} & \textcolor{black}{0.9802} & \textcolor{black}{0.9598} & \textcolor{black}{0.9908} & \textcolor{black}{0.9920}\\
\addlinespace[1ex]
scr$_{\text{var}}$ & \textcolor{black}{0.9795} & \textcolor{black}{0.9881} & \textcolor{black}{0.9829} & \textcolor{black}{0.9821} & \textcolor{black}{0.9835} & \textcolor{black}{0.9854} & \textcolor{black}{0.9701} & \textcolor{black}{0.9475} & \textcolor{black}{0.9770} & \textcolor{black}{0.9773}\\
scr$_{\text{cov}}$ & \textcolor{black}{0.9837} & \textcolor{black}{0.9902} & \textcolor{black}{0.9867} & \textcolor{black}{0.9869} & \textcolor{black}{0.9879} & \textcolor{black}{0.9884} & \textcolor{black}{0.9745} & \textcolor{black}{0.9547} & \textcolor{black}{0.9813} & \textcolor{black}{0.9816}\\
\addlinespace[1ex]
gw & \textcolor{black}{\textbf{0.9290}} & \textcolor{black}{\textbf{0.9283}} & \textcolor{black}{\textbf{0.9381}} & \textcolor{black}{\textbf{0.9405}} & \textcolor{black}{\textbf{0.9363}} & \textcolor{black}{\textbf{0.9390}} & \textcolor{black}{\textbf{0.9383}} & \textcolor{black}{\textbf{0.9380}} & \textcolor{black}{\textbf{0.9369}} & \textcolor{black}{\textbf{0.9359}}\\
\bottomrule
\end{tabular}